\documentclass[useAMS,usenatbib,usegraphicx,rotating,color,multirow,upgreek]{aa}  
\usepackage{graphicx}
\usepackage{txfonts}
\usepackage{epsfig}
\usepackage{amsmath} 
\usepackage{rotating}           
\usepackage{color}     
\usepackage{graphicx}
\usepackage{upgreek} 

\def\Mo{M$_\odot$}

\def\kms {\hbox{${\rm km\ s}^{-1}$}}
\def\ccm {$\hbox{{\rm cm}}^{-3}$}    
\def\scm  {$\hbox{{\rm cm}}^{-2}$}    
\def \AL {$\alpha $}     
\def \HI {H{\sc \,i}}
\def \HII {H{\sc \,ii}}

\def\lapp{\ifmmode\stackrel{<}{_{\sim}}\else$\stackrel{<}{_{\sim}}$\fi}
\def\gapp{\ifmmode\stackrel{>}{_{\sim}}\else$\stackrel{>}{_{\sim}}$\fi}

\begin{document} 
\title{High spin temperatures at large impact parameters: Ionisation in the outskirts of galaxies} \author{S. J. Curran}
\institute{School of Chemical and Physical Sciences, Victoria University of Wellington, PO Box 600, Wellington 6140, New Zealand\\
  \email{Stephen.Curran@vuw.ac.nz} 
} \abstract{ 
By including the most recent observations of \HI\ 21-cm absorption through nearby galactic discs, we
  confirm our previous assertion that there is an anti-correlation between the abundance of cool neutral atomic gas and
  impact parameter. Comparing the measured neutral hydrogen column densities of the sample with the absorption
  strength, we find a peak in the mean spin temperature of $<T_{\rm spin}/f>\approx2310$~K at an impact parameter of
  $\rho\approx14$~kpc,  with $<T_{\rm spin}/f>\gapp1000$~K in the remainder of the disc. This is significantly different to the spin
  temperature distribution in the Milky Way, which exhibits a constant $\approx250-400$~K over $\rho=8-25$~kpc.  The measured
  column densities may, however, suffer from beam dilution, which we show appears to be the case for the observations of
  \HI\ 21-cm emission in which the beam subtends radii of $\gapp10$~kpc. We therefore applied the column density profile
  of the Milky Way, in addition to the mean of the sample, observed at sufficiently high resolution, and the mean
  profile for the nearby $\sim10^{12}$\,\Mo\ galaxies in the IllustrisTNG simulations.  All of the models yield a peak
  in the mean spin temperature at similar impact parameters ($r\approx10-15$~kpc) as the measured column
  densities. These radii are similar to those of the spiral arms where \HII\ regions are often concentrated. We
  therefore suggest that the elevated spin temperatures trace the \HII\ regions observed in the outer disc of many spiral galaxies.
} 


 \maketitle
%

\section{Introduction} 
\label{intro}

Absorption of the 21-cm flux from distant radio sources through the discs of nearby galaxies provides information on
the neutral hydrogen (\HI) gas. In particular, a comparison of the absorption strength to the total \HI\ column density
yields the spin temperature of the gas, which is a measure of the population of the upper hyperfine level relative to
the lower level.  This can be increased via excitation by 21-cm absorption \citep{pf56}, excitation above the ground state
by Lyman-$\alpha$ absorption \citep{fie59} and collisional excitation \citep{be69}. The spin temperature gives the
fraction of the cold neutral medium (CNM, where $T\sim100$~K and $n\sim10$~\ccm) to the warm neutral medium (WNM, where
$T\sim2000$ K and $n\sim0.4$ \ccm, \citealt{sof18}).  Both the total gas density (e.g. \citealt{too63}) and fraction of
cool neutral gas (\citealt{cras16}) exhibit a decrease in abundance with the galactocentric radius. The latter, based on the
then 90 sight-lines searched for \HI\ 21-cm absorption, has, however, since been refuted by subsequent studies
\citep{bor16,dsg+16}. A flat 21-cm absorption strength in conjunction with a decreasing column density would result in
temperature climb outwards across the disc. This runs contrary to what is seen in the Milky Way, where the mean spin
temperature maintains $<T_{\rm spin}>=250 - 400$~K over radii of 8 -- 25~kpc \citep{dsg+09}.

From the decrease in 21-cm absorption strength with the impact parameter, \citet{cras16} find the mean spin temperature
to peak, with $<T_{\rm spin}>\approx2900$~K, at $\rho\approx10$~kpc.  This, however, was based on a sample of just 27
sight-lines with measured column densities. Additionally, given that spin temperatures of $\approx500$~K and
$\approx200$~K were found for the inner and outer disc, respectively, this was deemed to be consistent with the Galactic
values.  With these new data, the number of sight-lines is now 143, for which 39 have column density measurements.  Here
we add the new data to the previous, confirming that the 21-cm absorption strength is anti-correlated with impact
parameter, in addition to using the column densities to determine the mean spin temperature of the gas across the disc.

\section{Analysis}
\subsection{Absorption strength versus impact parameter} 

The addition of subsequent data \citep{bor16,dsg+16,asd+19} to that used in \citet{cras16}\footnote{Compiled from
  \citet{hb75,bdkb88,cs90,cv92,kac02,hc04,bty+10,bty+11,bmh+14,gsb+10,gsn+13,sgr+13,rsa+15,rsa+16,zlp+15,dgso16}.},
brings the number of sight-lines from 90 to 143. This comprises 22 detections of 21-cm absorption and 121 upper
limits. In order to include these, we re-sample the limits to a common spectral resolution/profile width  of 10 \kms\ (see
\citealt{cur12}) and then flag these as censored data points using the {\em Astronomy SURVival Analysis} ({\sc asurv})
package \citep{ifn86}.

\begin{figure}
\centering \includegraphics[angle=-90,scale=0.54]{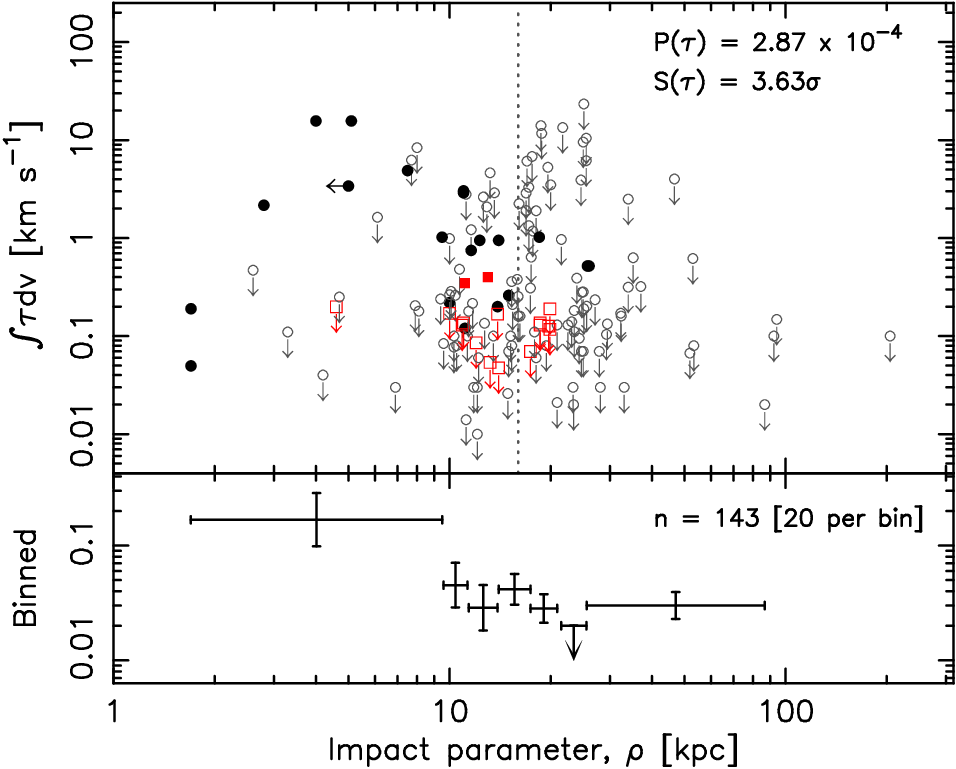}
\caption{Velocity integrated optical depth versus the impact parameter for all of the published searches. 
The circles represent the previous data (\citealt{cras16} and references therein) and the squares the data since
\citep{bor16,dsg+16,asd+19}.The downward arrows signify the $3\sigma$ upper limits 
and the dotted vertical line shows the median impact parameter.  The bottom panel shows the binned values,
  including the limits via the Kaplan--Meier estimator, in equally sized bins. The horizontal error bars show the range
  of points in the bin and the vertical error bars the $1\sigma$ uncertainty in the mean value.}
\label{impact}
\end{figure}

Plotting the integrated optical depth versus the impact parameter  (Fig. \ref{impact}),
a generalised non-parametric Kendall-tau test gives a probability of $P(\tau) = 2.87\times10^{-4}$ of the observed $\int\tau dv$--$\rho$ anti-correlation  arising by chance. Assuming Gaussian statistics, this 
is significant at $S(\tau) = 3.63\sigma$, cf. $3.31\sigma$ previously \citep{cras16}. 
Neither \citet{bor16} nor \citet{dsg+16}  find a strong correlation, although the former only test their sample 
of one 21-cm absorption detection and 15 non-detections (all at $\rho<20$~kpc) and the latter 
limit their impact parameters to $\rho<30$~kpc, resulting in 
$S(\tau) = 2.42\sigma$. 

The $\int\tau dv$--$\rho$ dependence is confirmed via the detection rates above and below the median impact parameter of
$\rho= 16$ kpc.  At $\rho \leq 16$ kpc there are 19 detections and 53 non-detections, giving 26.4 per-cent detection
rate. At $\rho>16$ kpc there are three detections and 68 non-detections. Based on a likelihood of $p= 0.264$, the
binomial probability of obtaining three detections or fewer out of 71 sight-lines is $6.16\times10^{-6}$, which is
significant at $4.50\sigma$.

\subsection{Spin temperature versus impact parameter}

Several of the sight-lines have been detected in 21-cm emission from which the column 
density can be obtained from the brightness temperature of the line emission, $T_{\rm b}$~ [K].
In the optically thin regime ($\tau\lapp0.3$) 
the  column density, $N_{\text{\HI}}$~[\scm]  is given by
\begin{equation}
N_{\text{\HI}}  = 1.823\times10^{18} \int\! T_{\rm b}\,dv,
\label{em_eq}
\end{equation}
and in absorption the column density is related to the spin temperature, $T_{\rm spin}$ [K], via
\begin{equation}
N_{\text{\HI}}  =1.823\times10^{18}\,\frac{T_{\rm  spin}}{f}\int\!\tau\,dv,
\label{N_eq}
\end{equation}
where $f$ is the covering factor; the fraction of the background flux intercepted by the absorbing gas. Thus, the
comparison of the velocity integrated optical depth with the column density yields the spin temperature degenerate with
the covering factor, $T_{\rm spin}/f$. For unresolved galaxies detected in 21-cm absorption at non-zero redshifts, 
the covering factor is dependent upon the relative
absorber and emitter cross-sections as well as the redshift of the background source \citep{cur12}. Given, that here the
absorption is occuring through a nearby resolved disc, it is fair to assume that all of the background flux is
intercepted, giving $f\approx1$. 

\subsubsection{Measured column densities}

For 39 of the sight-lines 21-cm emission has been detected, giving the \HI\ column density. Where this is 
not given (e.g. \citealt{bor16}),
we obtain the brightness temperature from the \HI\ mass, $M_{\text{\HI}}$, via the integrated flux density of the emission (e.g. \citealt{rw00})
\begin{equation}
M_{\text{\HI}} = 236 D_{\rm L}^2S_{\rm int}, \text{ where }  S_{\rm int}  = \frac{2k\Omega_{\rm b}} {\lambda^2}\int\! T_{\rm b}\,dv,
\end{equation}
$D_{\rm L}$ is the luminosity distance  [Mpc], $S_{\rm int}$  the integrated flux of the line [mJy~\kms], 
$k$ the Boltzmann constant, $\Omega_{\rm b}$ the beam solid angle of the telescope at the wavelength $\lambda$ [m]
and  $\int\! T_{\rm b}\,dv$  the velocity integrated brightness temperature [K~\kms]. 

\begin{figure}
\centering \includegraphics[angle=-90,scale=0.54]{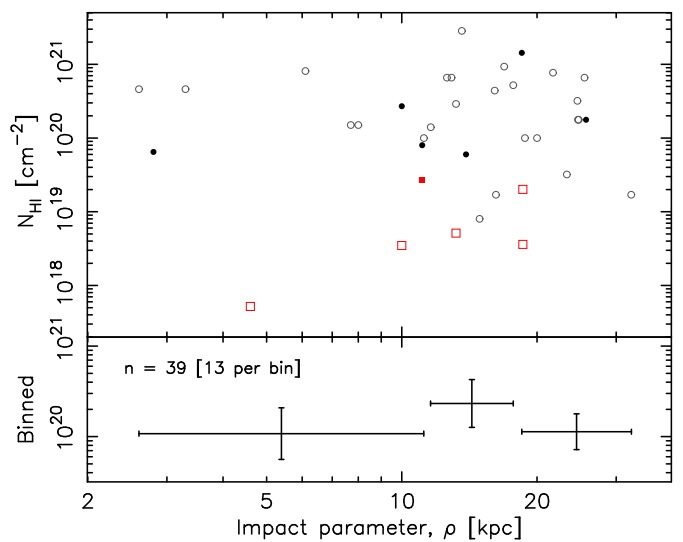}
\caption{Column density versus the impact parameter for the sight-lines detected in 21-cm emission.} 
\label{N-impact}
\end{figure}

\begin{figure}
\centering \includegraphics[angle=-90,scale=0.54]{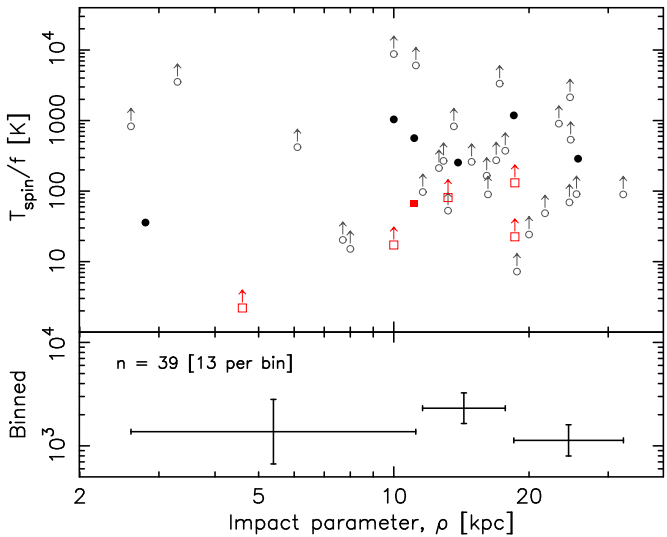}
\caption{Spin temperature/covering factor obtained from the column density (Fig.~\ref{N-impact}),
versus the impact parameter. The upward arrows signify lower limits.} 
\label{spin-impact}
\end{figure}

In Fig.~\ref{N-impact} we show the distribution of column density with impact parameter
for the sight-lines with detected 21-cm emission. From this, we see that the column density is fairly consistent out to
$\rho\approx25$~kpc, with a possible elevation at $\approx15$~kpc. 
In Fig.~\ref{spin-impact}, we show the resulting spin temperatures, which 
exhibits a similar $T_{\rm spin}/f$ bump as previously \citep{cras16}, the binning of the data giving $<T_{\rm
  spin}/f>=1370^{+1440}_{-700}$~K at $\rho\approx5$~kpc, $<T_{\rm spin}/f>=2310^{+940}_{-670}$~K at $\rho\approx14$~kpc
and $<T_{\rm spin}/f>=1100^{+470}_{-220}$~K at $\rho\approx25$~kpc.
These temperatures are considerably higher than
in the Milky Way (Sect.~\ref{intro}). 

\begin{figure}
\centering \includegraphics[angle=-90,scale=0.50]{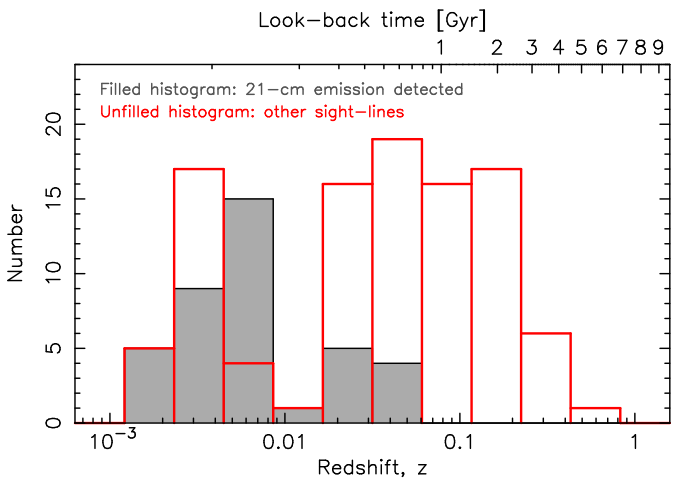}
\caption{Redshift distribution of the target galaxies. 
The galaxies span the redshift range  $0\lapp z \leq 0.4367$, with 21-cm emission being detected to 
$z = 0.0462$ (filled histogram).} 
\label{z-hist}
\end{figure}

\begin{figure}
\centering \includegraphics[angle=-90,scale=0.54]{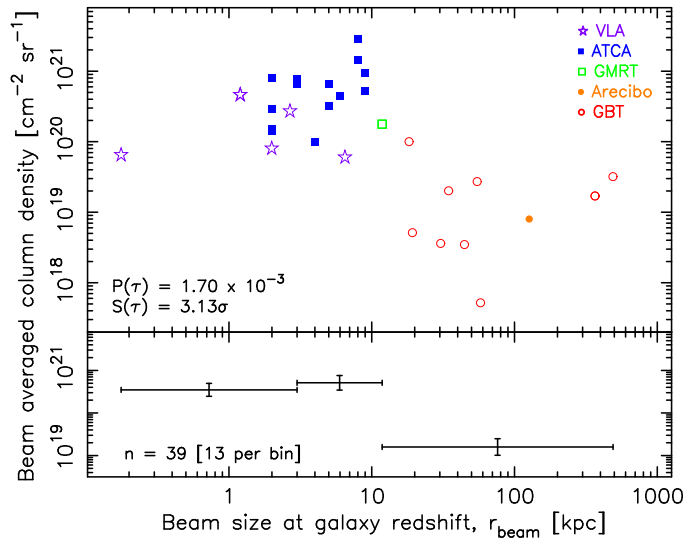}
\caption{Column density measured across the telescope beam versus the linear extent of the beam for the sight-lines
  detected in 21-cm emission. The telescopes used are the Very Large Array (VLA, \citealt{cv92,bty+11,bmh+14}), the
  Australia Telescope Compact Array (ATCA, \citealt{rsa+15,rsa+16}), the Giant Metre-Wave Radio Telescope (GMRT,
  \citealt{dgso16}), the Arecibo telescope \citep{cs90} and the Green Bank Telescope (GBT,
  \citealt{hb75,bty+11,bor16}).}
\label{N-beam}
\end{figure}

Determining the spin temperature through the comparison of Equs. \ref{em_eq} and \ref{N_eq} relies
upon the same sight-line being probed, which may not be the case here: In absorption we observe
a `pencil-beam' through the gas, whereas for emission we observe over the whole telescope beam, which results in
dilution if the beam size exceeds that of the emitting gas ($r_{\rm beam} > r_{\rm gas}$). 
The linear extent of the beam is given by
$ r_{\rm beam} = \theta_{\rm beam}D_{\rm A}$,
where  $\theta_{\rm beam}$ is the angular beam size 
[radians] and $D_{\rm A}$ the angular diameter distance (e.g. \citealt{pea99}) to the galaxy [kpc], obtained from the redshift
(Fig.~\ref{z-hist}).
Showing $N_{\text{\HI}}$ versus $r_{\rm beam}$ in Fig.~\ref{N-beam}, we see a clear dependence between the measured column
density and the beam width.

\subsubsection{Galactic column density}
\label{gcd}

\begin{figure}
\centering \includegraphics[angle=-90,scale=0.49]{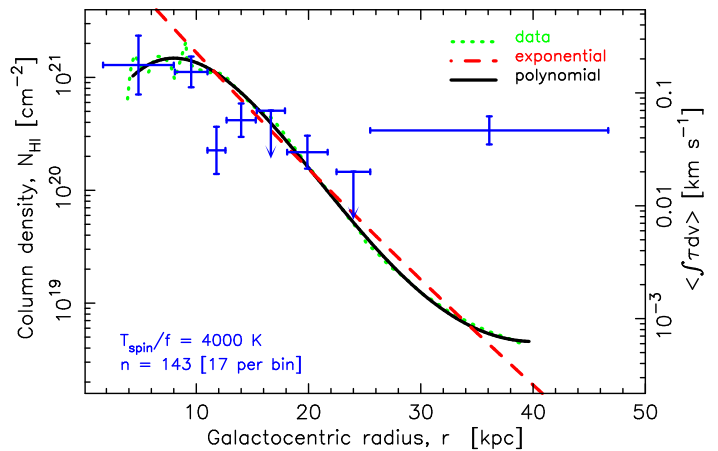}
\caption{Radial column density distribution of the Milky Way (total disc, \citealt{kd08}) overlain with
the exponential fit to the $r\geq12.5$~kpc data \citep{kk09} and a 
log-polynomial to all of the data. 
The error bars show the  column density of the 21-cm absorption  for $T_{\rm spin}/f = 4000$~K (right axis).}
\label{gal}
\end{figure}

In order to bypass the effect of beam dilution on the column density, we can apply well constrained profiles 
in similar galaxies to our sample.
One option is to use the Milky Way, where the total (north and  south, \citealt{kd08}) disc distribution is shown in Fig.~\ref{gal}. 
We overlay
the mean integrated optical depths of the 21-cm absorption for a constant spin temperature of 
$T_{\rm spin}/f \approx4000$~K,  which traces the column density out
to $r\approx20$~kpc, beyond which  the spin temperature decreases due to the edge of the stellar disc being reached.
These spin temperatures are much higher than observed in the Milky Way \citep{dsg+09} and,
unlike the Galactic values, appear to be further elevated at $r\approx10$~kpc.

\begin{figure}
\centering \includegraphics[angle=-90,scale=0.51]{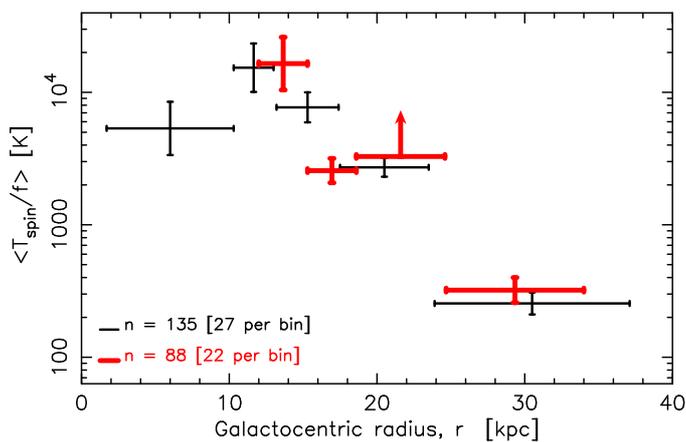}
\caption{Spin temperature/covering factor values obtained from the polynomial (thin bars) and exponential (thick bars)
 fits to the Galactic  column density distribution (Fig. \ref{gal}).}
\label{gal_spin}
\end{figure}

Using both the exponential fit (for $12.5\leq r\leq35$~kpc, \citealt{kk09}) and the
log-polynomial fit (for $r\leq40$~kpc), which provides a better trace of the inner disc, we show the
spin temperature profile in Fig.~\ref{gal_spin}. Where the radii overlap, these give 
similar results and, again, 
a peak in the spin temperature of $T_{\rm spin}/f\approx15\,000$~K at $r\approx10-15$~kpc, with
the  inner and outer stellar disc having temperatures $T_{\rm spin}/f \approx3000-5000$~K (see Table~\ref{summ}).

\begin{figure}
\centering \includegraphics[angle=-90,scale=0.51]{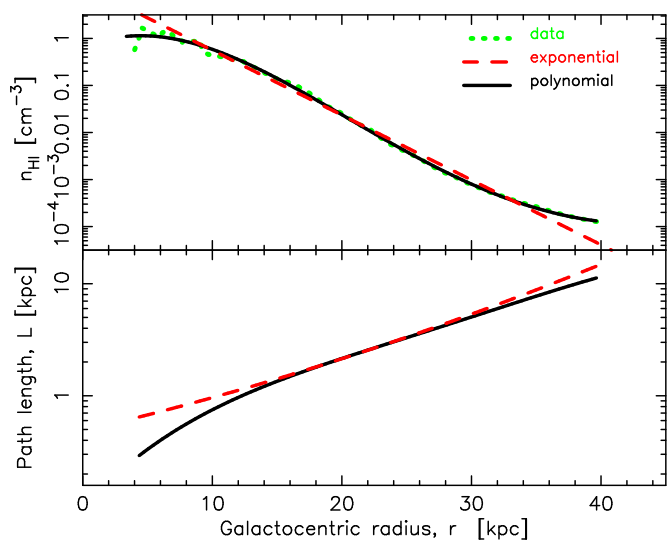}
\caption{Top: Volume density profile of the Milky Way  (total disc, \citealt{kd08}), overlain with the exponential 
and a log-polynomial fit. Bottom: The resulting path length profile.}
\label{path}
\end{figure}

\begin{figure}
\centering \includegraphics[angle=-90,scale=0.51]{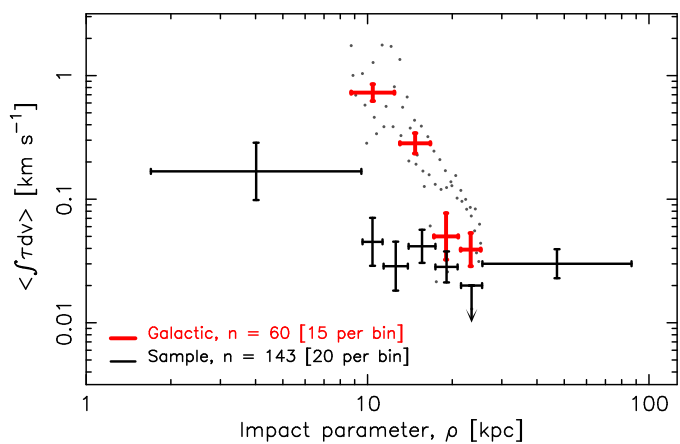}
\caption{Binned mean velocity integrated optical depth of the sample (thin bars) in comparison to the Milky Way:  The data
from each of the Canadian Galactic Plane Survey (CGPS, \citealt{tgp+03}), the
  Southern Galactic Plane Survey (SGPS, \citealt{mdg+05}) and the VLA Galactic Plane Survey (VGPS, \citealt{std+06})
are shown as points,  with the thick bars showing the binned values.}
\label{dsg}
\end{figure}

Dispensing with the assumed column density profile, we can compare the mean 21-cm absorption strengths (Fig.~\ref{impact})
directly with those of the Milky Way. \citet{dsg+09} give these in terms of $\int\tau d. /L$, where the path length $L =
N_{\text{\HI}}/n_{\text{\HI}}$ and $n_{\text{\HI}}$ is the volume density of the gas. From the column density
(Fig.~\ref{gal}) and mid-pane volume density profiles \citep{kd08} of the Milky Way, we obtain the path length
distribution shown in Fig. \ref{path},
which exhibits the flaring of the scale-height of the \HI\ disc \citep{kdkh07}. Showing the resulting
integrated optical depths in Fig.~\ref{dsg},
we see that the mean sample strengths are significantly weaker than the Galactic values, especially in the inner disc.
For a given
column density profile this will result in higher spin temperatures for the sample, thus indicating
that a very different profile is required to yield temperatures similar to that in the Milky Way.

\subsubsection{Mean sample column density}
\label{mcd}

\begin{figure}
\centering \includegraphics[angle=-90,scale=0.52]{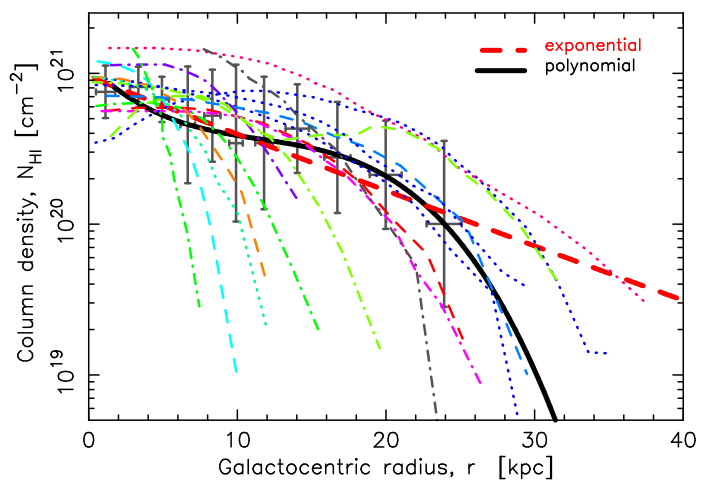}
\caption{\HI\ column density profiles from the high resolution observations of \citet{rsa+15,rsa+16}. The
error bars show the mean values and the $\pm1\sigma$ uncertainties. The thick unbroken curve
shows the  ($1/\sigma$) weighted log-polynomial fit and the broken line the weighted exponential fit to
the mean values.} 
\label{rsa}
\end{figure}

\begin{figure}
\centering \includegraphics[angle=-90,scale=0.50]{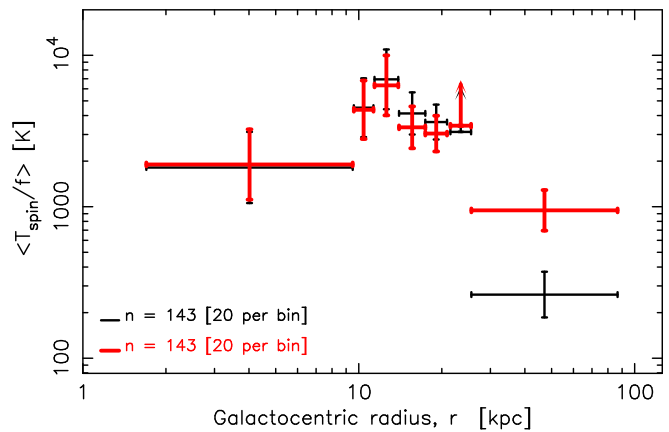}
\caption{As per Fig.~\ref{gal_spin} but for the mean sample column density distribution (Fig. \ref{rsa}).} 
\label{rsa_spin}
\end{figure}

Another option for the column density profile is to use those of the sample itself, which we show 
in Fig.~\ref{rsa}, where available.\footnote{From
high resolution observations with the ATCA (\citealt{rsa+15,rsa+16}).}
The profiles are very diverse and so we model the column density distribution via a weighted 
log-polynomial and a weighted exponential fit to the mean values. These show reasonable agreement out to
$r\approx20$~kpc, beyond which the polynomial appears to provide the best mean trace. In Fig.~\ref{rsa_spin}, we show
the spin temperature profile generated by both fits.
Again, these are higher than in the Milky Way, but lower than that obtained using the Galactic column density distribution.
There is, however, a similar peak in the spin temperature, $T_{\rm spin}/f =7000^{+4000}_{-2500}$~K
at $r\approx12$~kpc, with  $T_{\rm spin}/f \approx2000-3000$~K in the remainder of the stellar disc.

\begin{figure}
\centering \includegraphics[angle=-90,scale=0.49]{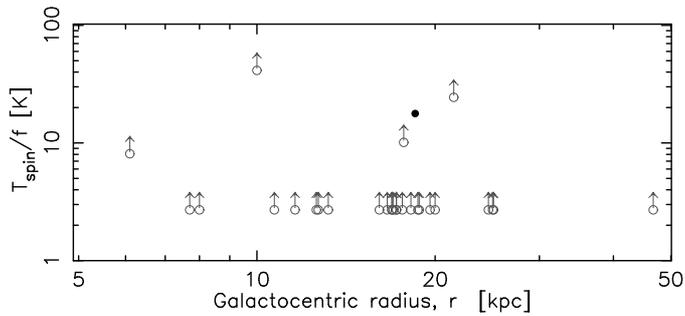}
\caption{Spin temperature/covering factor values at various impact parameters for the 21-cm observations and column
density profiles of \citet{rsa+15,rsa+16}. The symbols are per Fig.~\ref{spin-impact}, where the limits with $T_{\rm spin}/f < T_{\rm CMB}$ have been capped at $T_{\rm spin}/f \geq 2.7$~K.}
\label{rsa_ind}
\end{figure}

We acknowledge that this is based upon the mean column density distribution, where the individual cases 
differ drastically. Ideally, we would determine  the spin temperature from the 21-cm absorption strength scaled
by the relevant column density for each individual sight-line. Doing this for the high resolution observations (Fig.~\ref{rsa_ind}),
we see, however, that due, partly at least, to the high impact parameters probed, this yields almost exclusive
limits, most of which are relatively weak.
Thus, we have little choice but to use the averaged 21-cm absorption strengths (Fig.~\ref{impact}), normalised by a mean 
column density distribution in order to yield a (mean) spin temperature profile.

\subsubsection{Simulated column density}
\label{scd}

\begin{figure}
\centering \includegraphics[angle=-90,scale=0.49]{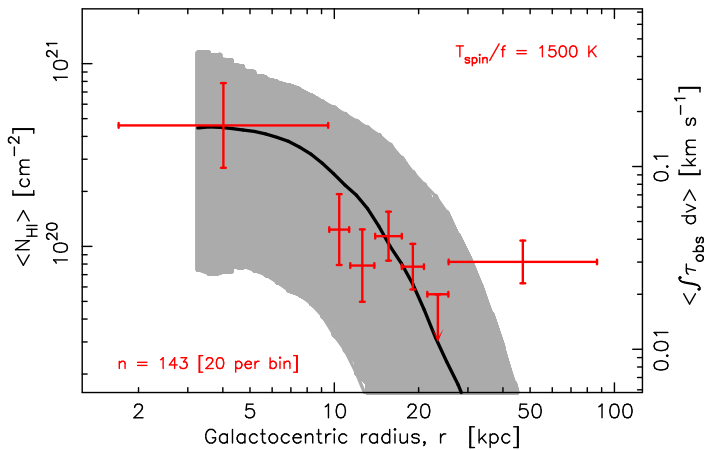}
\caption{Mean \HI\ column density from an ensemble of $z=0$ galaxies from the TNG (left axis), with the
  integrated column density of the 21-cm absorption overlain (for $T_{\rm spin}/f = 1500$~K, right axis). The shaded
  region shows the 16th and 84th percentiles ($1\sigma$) of the population variation in the TNG. }
\label{TNG}
\end{figure}

\begin{figure}
\centering \includegraphics[angle=-90,scale=0.49]{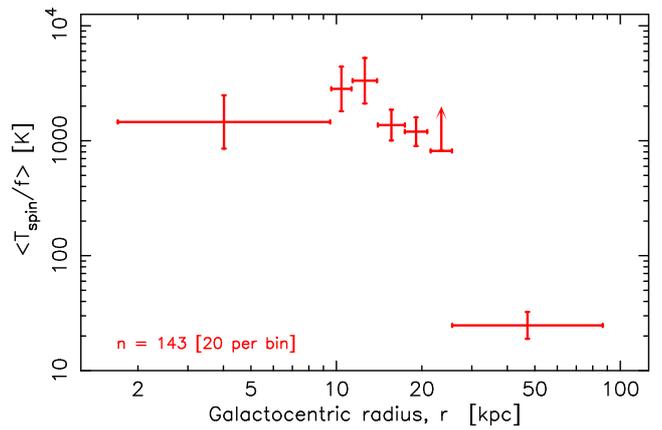}
\caption{Spin temperature/covering factor values obtained from the mean TNG column density profile (Fig.~\ref{TNG}).
}
\label{spin}
\end{figure}

Both the column density profiles of the Milky Way and the mean sample yield spin temperatures much higher than Galactic
values.  This could be due to the relatively high column densities, with $N_{\text{\HI}}\sim10^{21}$~\scm\ at
$r\approx0$ in both cases. Lower column densities can be obtained from the mean profile generated by the IllustrisTNG
simulations for the 567 $z=0$ galaxies with $\log_{10}M = 11.8-11.9$\,\Mo\ (\citealt{nsp+19}, Fig.~\ref{TNG}).
Comparing the 21-cm absorption strength with the column density distribution, we find, as for the 
Galactic distribution (Fig.~\ref{gal}), that the profile is well fit for a constant $T_{\rm spin}/f$, asides from a 
peak at $r\approx12$~kpc and a drop in spin temperature beyond the stellar disc. However, the spin
temperature required is much lower than the $T_{\rm spin}/f \approx4000$~K from the Galactic $N_{\text{\HI}}$ 
distribution, with the profile (Fig.~\ref{spin}) giving $T_{\rm spin}/f \approx1200- 1500$~K in the stellar disc,
 with the $r\approx12$~kpc peak persisting (with $T_{\rm spin}/f =3300^{+1900}_{-1200}$~K, Fig.~\ref{spin}).

\section{Discussion}
\label{sec:disc}

All of the above models (summarised in Table~\ref{summ}) lead to significantly higher spin temperatures than in 
\begin{table*}
 \caption{Values of $<T_{\rm spin}/f>$~[K] at various impact parameters for the various column density profiles (Sects. \ref{gcd} - \ref{scd}). }
\centering
\bgroup
\def\arraystretch{1.5}
\begin{tabular}{l cc c cccc} 
\hline\hline
\smallskip  
Model &   Fig. & \multicolumn{5}{c}{Approximate radius/impact parameter  [kpc]}\\

                &    & 5 & 12  & 15 & 20 & 30 \\
\hline
Measured          &  \ref{spin-impact}     & $1400^{+1400}_{-700}$ & --- & $2300^{+900}_{-700}$  & \multicolumn{2}{c}{$1100^{+500}_{-300}$  at 25~kpc}\\
Galactic -- poly &  \ref{gal_spin}     & $5300^{+3100}_{-2000}$ & $15\,000^{+8000}_{-5000}$ & $7700^{+2300}_{-1800}$  & $2700^{+500}_{-400}$ & $250^{+50}_{-40}$\\
~~~-- exponential &  & ---& $16\,000^{+10000}_{-6000}$ & $2600^{+600}_{-500}$  & $\geq3300$ & $320^{+80}_{-60}$\\
Sample -- poly&  \ref{rsa_spin}     & $1800^{+1300}_{800}$ & $6900^{+4000}_{-2500}$     & $4100^{+1600}_{-1100}$     & $3600^{+1100}_{-800}$  & $460^{+110}_{-90}$\\
~~~-- exponential & & $1900^{+1300}_{800}$ & $6300^{+3700}_{-2300}$     & $3300^{+1300}_{-900}$& $3000^{+900}_{-700}$& $1100^{+270}_{-220}$\\
Simulated  &   \ref{spin}    & $1500^{+1000}_{600}$ &  $3300^{+1900}_{-1200}$ & $1400^{+500}_{-400}$     & $1200^{+400}_{-300}$  & $180^{+40}_{-30}$\\
\hline
\end{tabular} 
\egroup
\label{summ}  
\end{table*}
the Milky Way, with each exhibiting a peak at $r\approx12$~kpc. Beyond this, the decrease in spin temperature
is most likely due to the \HI\ disc extending  beyond the stellar disc (e.g. \citealt{ckb08,wbd+08}). However,
temperatures similar to that within the inner Milky Way are not reached until $r\approx30$~kpc.
The high $T_{\rm spin}/f$ values could be accounted for
by covering factors $f\sim0.1$, although we expect $f\sim1$ for the absorption of unresolved quasar emission through a resolved disc.  

Regarding the peak, in spiral galaxies, including the Milky Way \citep{loc76}, the ionised gas (\HII) regions are
concentrated in the spiral arms, at radii of up to $r\approx15$~kpc (e.g. \citealt{hod69,hk83,jbk09}), believed to be
caused by OB stars \citep{mwc53}.\footnote{Although it has been argued that the arms concentrate the gas without
  triggering star formation \citep{frw+10}.  Furthermore, while \citet{sj02} find a correlation between spiral arm
  strength and star formation, \citet{frw+10} and \citet{emm+13} find the inter-arm star formation rates to be similar
  to those in the arms.  Compounding the issue further is the fact that the star formation rate is correlated with the
  molecular gas abundance, as traced through CO \citep{blw+08,lws+13}, although the spiral arms to do not appear to host
  elevated CO levels \citep{ee87,dsg+09,ksh16}.}  The ionising ($\lambda\leq912$~\AA) photon rate from each star is
given by $Q_{\star}\equiv\int^{\infty}_{\nu_0}({L_{\nu}}/{h\nu})\,d{\nu}$, where $\nu_0 = 3.29\times10^{15}$~Hz,
$L_{\nu}$ is the specific luminosity at frequency $\nu$ and $h$ the Planck constant.  For a radiative recombination rate
coefficient of $\alpha_{\rm B}$ and proton and electron volume densities $n_{\rm p}$ and $n_{\rm e}$, respectively, the
equilibrium between the photo-ionisation and recombination of protons and electrons is \citep{ost89}
\begin{equation}
Q_{\star} = 4\pi\int^{r_{\rm s}}_{0}\,n_{\rm p}\,n_{\rm e}\,\alpha_{\rm B}\,r^2\, dr  \Rightarrow  Q_{\star} =\frac{4}{3}\pi r_{\rm s}^3 n_{\text{\HI}}^2 \alpha_{\rm B},
\label{eq1}
\end{equation}
for a neutral plasma, $n_{\rm p} = n_{\rm e} = n_{\text{\HI}}$ of constant density, 
defining  the Str\"{o}mgren sphere.

For a star of effective temperature $T_{\star}$, we obtain the ionsing photon rate 
from the integrated intensity of the Planck function, $I_{\nu}$, 
\begin{equation}
Q_{\star}= 4\pi R_{\star}^2\int^{\infty}_{\nu_0}\frac{I_{\nu}}{h\nu}d{\nu},
\text{ where } I_{\nu}= \frac{2\pi h\nu^3}{c^2} \frac{1}{e^{h\nu/kT_{\star}}-1}
\label{Q}
\end{equation}
and $4\pi R_{\star}^2$ is the surface area of the star. We estimate this by comparing the bolometric intensity, 
$I= \int^{\infty}_{0}I_{\nu}d{\nu}$, with the bolometric luminosity obtained from the main sequence
($\log_{10}L_{\rm MS} = 6.50 \log_{10}T_{\star} -24.37$), giving  the values of $Q_{\star}$ listed in Table~\ref{tab:Q}.
\begin{table} 
  \caption{The estimated properties of stars of various  effective temperatures, $T_{\star}$. 
    $L_{\rm MS}$ is the bolometric luminosity estimate from the temperature of a main sequence star, followed 
    by the radius required to equate this to the bolometric intensity and finally
    the number of ionising photons per second.}
\begin{tabular}{l  c  c c c} 
\hline\hline
\smallskip
$T_{\star}$ [K] & $L_{\rm MS}$ [L$_{\odot}$] & Radius [R$_{\odot}$] &$Q_{\star}$ [s$^{-1}$]  \\
\hline
10\,000  & 46    & 2.3 & $7.8\times10^{41}$\\
 20\,000 & 4200  & 5.4 & $2.7\times10^{46}$\\
 30\,000  &59\,000 & 9.0 & $1.8\times10^{48}$\\
 40\,000  & $4\times10^5$ &13 & $2.0\times10^{49}$\\
 50\,000 & $2\times10^6$ &17 & $1.1\times10^{50}$\\ 
\hline
\end{tabular}
\label{tab:Q}  
\end{table} 

\begin{figure}
\centering \includegraphics[angle=-90,scale=0.50]{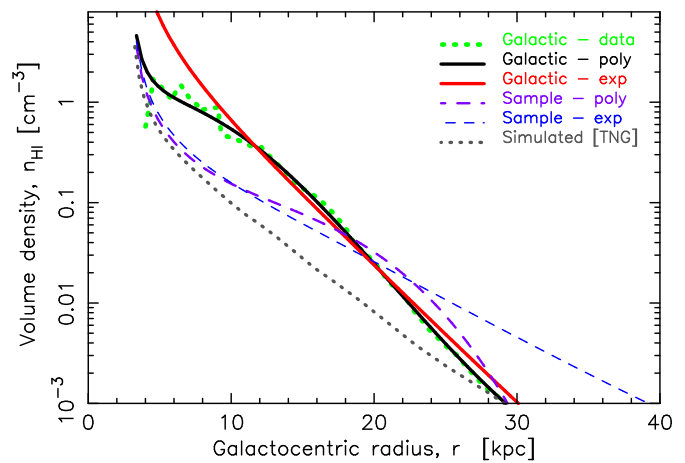}
\caption{Estimated volume densities for the various column density profiles obtained from the Galactic
log-polynomial path length fit  (Fig.~\ref{path}).}
\label{vol}
\end{figure}

\begin{figure}
\centering \includegraphics[angle=-90,scale=0.52]{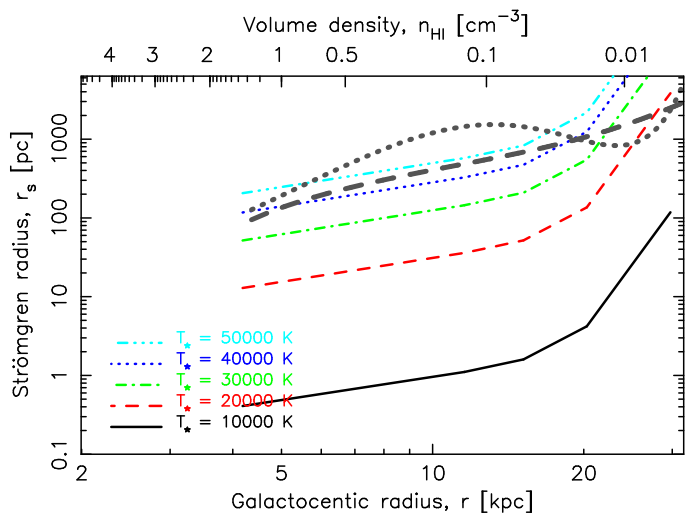}
\caption{Str\"{o}mgren radius versus the Galactocentric radius for stars of different effective temperatures
obtained from the log-polynomial volume density fit to the sample (Fig.~\ref{vol}).
The thick grey dotted curve shows the half-path length, $L/2$ (where $L = N_{\text{\HI}}/n_{\text{\HI}}$), for the mean of the sample and the dashed line  for the Milky Way (Fig.~\ref{path}).}
\label{strom}
\end{figure}

Applying these to Equ.~\ref{eq1}  yields the Str\"{o}mgren radius  of each star.
For the radiative recombination rate coefficient, $\alpha_{\rm B}$, we assume that the electron temperature is
equivalent to the spin temperature ($T_{\rm e} = T_{\rm spin}$).
Arising from different heating/cooling processes, this is may not be justified although,
given that $\alpha\propto\sqrt{T_{\rm e}}$,
the temperature dependence is not strong.\footnote{http://amdpp.phys.strath.ac.uk/tamoc/DATA/RR/} 
From Equ.~\ref{eq1} we see that the Str\"{o}mgren  radius is dominated by the volume density, which we
obtain at a given Galactocentric radius by  assuming the log-polynomial $L-r$ profile of the sample (Fig.~\ref{vol}).
These give the  Str\"{o}mgren radii shown in Fig.~\ref{strom}.
Due to the decrease in gas density, for a homogeneous, dust-free medium, we see that the size of the sphere
increases significantly with galactocentric radius. 

In disc galaxies the star formation rate
is found to increase with galactocentric radius \citep{mgb+07,abt09}. This could be the result
of the denser environments closer to the galactic centre suppressing the star formation \citep{bgb+07,wch+08}. 
The outward growth of galactic discs would mean that the large radii are dominated by the younger 
(and more short-lived) stellar populations \citep{tp05,mgb+07}, giving a larger fraction of the more luminous
stars (Table.~\ref{summ}).
For the O-type stars ($T_{\star}\gapp30\,000$~K), we see that the Str\"{o}mgren radius for an individual star
is comparable to the width of the \HI\ disc at $r\gapp20$~kpc. Given the approximations, the values
are uncertain, although it is apparent that, due to the decreasing density of the surrounding medium, 
the Str\"{o}mgren radii exhibit a steep increase, which is consistent with the gas being highly ionised 
at large impact parameters.

\section{Conclusions}

From the absorption of the 21-cm continuum from 90 sight-lines towards distant radio sources through the discs of
nearby spiral galaxies, \citet{cras16} reported an anti-correlation between the abundance of cool, star-forming, gas
and the galactocentric radius. Since the abundance of this gas normalised by the total \HI\ column density gives a
measure of the spin temperature, such an anti-correlation would be expected if the mean spin temperature where constant
across the disc, as it is in the Milky Way (out to $r\approx20$~kpc, \citealt{dsg+09}). The correlation has, however,
been disputed by subsequent data (increasing the number of sight-lines to 143, \citealt{bor16,dsg+16}).\footnote{With
  one additional sight-line from \citet{asd+19}.}  Including the new data in our analysis, we find the $\int\tau
dv$--$\rho$ anti-correlation to increase in significance to $S(\tau) = 3.63\sigma$ from $3.31\sigma$ \citep{cras16},
thus demonstrating that, like all of the neutral atomic gas, the cold component decreases in abundance with
galactocentric radius.

Comparing the mean 21-cm absorption strengths with the measured column densities, we find a possible peak in the spin
temperature of $<T_{\rm spin}/f>\approx2300$~K at $\rho\approx14$~kpc, compared to $<T_{\rm spin}/f>\lapp1400$~K in the
remainder of the stellar disc. We show, however, that the measured column densities obtained from the 21-cm emission
are likely to suffer significant dilution due to the beam subtending beyond the disc. Applying
better constrained column density profiles, we find:
\begin{itemize}
\item For the Galactic distribution, a peak of $<T_{\rm spin}/f>\approx15\,000$~K at $\rho\approx12$~kpc, with
$<T_{\rm spin}/f>\gapp3000$~K  in the remainder of the disc.
\item For the mean profile of the sample galaxies, where sufficiently high resolution data are available, 
a peak of $<T_{\rm spin}/f> \approx7000$~K at $\rho\approx12$~kpc, with
$<T_{\rm spin}/f>\gapp2000$~K  in the remainder of the disc.
\item For the mean of  a simulated ensemble of spiral galaxies, a peak of $<T_{\rm spin}>\approx3300$~K at $r\approx12$~kpc, with
$<T_{\rm spin}/f>\gapp1000$~K  in the remainder of the disc.
\end{itemize}
All of these spin temperatures are considerably higher than observed in the Milky Way ($T_{\rm spin}>\approx250-400$~K,
\citealt{dsg+09}), being closer to those observed in other low redshift galaxies detected through the absorption of
background quasar light, namely damped Lyman-\AL\ absorption systems (DLAs), where $T_{\rm spin}\gapp~1000$~K at
$z\approx0$. At the peak of the star formation history at $z\sim2$, however, the mean spin temperatures of the DLAs
approach those in the Milky Way \citep{cur19}.

We speculate that the elevated gas temperature at these radii may be coincident with the regions of highly ionised gas
observed in some nearby spirals. At $r\gapp10$~kpc, where  $n_{\text{\HI}}\lapp0.1$~\ccm, the radius of the \HII\
region around each O-type star exceeds $r_{\rm s}\sim100$~pc, which is a significant fraction of the path length
through the gaseous disc. Hence the presence of hot stars, in conjunction with the low gas densities, does indeed
suggest that the gas is highly ionised at large Galactocentric radii.
 
\section*{Acknowledgements}

I would like to thank the anonymous referee for their helpful feedback and  Dylan Nelson for the IllustrisTNG data.
This research has made use of the NASA/IPAC Extragalactic
Database (NED) which is operated by the Jet Propulsion Laboratory, California Institute of Technology, under contract
with the National Aeronautics and Space Administration and NASA's Astrophysics Data System Bibliographic Service. This
research has also made use of NASA's Astrophysics Data System Bibliographic Service and {\sc asurv} Rev 1.2
\citep{lif92a}, which implements the methods presented in \citet{ifn86}.


\label{lastpage}

\end{document}